\documentclass[fleqn,12pt,twoside]{article}
\usepackage[headings]{espcrc1}

\readRCS
$Id: espcrc1.tex,v 1.2 2004/02/24 11:22:11 spepping Exp $
\usepackage{graphicx}

\title{Highlights from the PHENIX Experiment - Part 2}

\author{H. B\"usching\address[BNL]{Brookhaven National Laboratory, Upton, NY 11973-5000, USA}
  (for the PHENIX\thanks{For the full list of PHENIX authors and acknowledgements,
    see appendix 'Collaborations' of this volume.}
Collaboration)}

\runtitle{Highlights from the PHENIX experiment - Part 2}
\runauthor{H. B\"usching}

\begin{document}

\maketitle

\begin{abstract}
Recent results from the PHENIX experiment at RHIC are presented.
In this second part of the overview we discuss new results on 
direct photon production at low $p_T$ in Au+Au collisions and
a systematic study of J/$\psi$ production in various systems.
Finally we summarize new results on the modification of jets
in the medium.
\end{abstract}


\section{PHOTONS AT LOW $p_T$}
Recently PHENIX published measurements of direct photons in Au+Au collisions at $\sqrt{s_{NN}} = 200$ GeV up 
to $p_T = 14$ GeV/c by comparing the measured inclusive photon $p_T$ spectrum with the expected 
spectrum of background photons from hadronic decays~\cite{Adler:2005ig}. A significant direct photon signal
could be measured for $p_T > 4$ GeV/c. In addition it could be shown that the magnitude of the signal
increases with increasing centrality. This is due to the decreasing decay background caused by $\pi^0$ suppression. 
In this $p_T$ range the direct photon measurement agrees very well with a binary scaled pQCD calculation~\cite{Jager:2002xm},
demonstrating that direct photons in Au+Au collisions at $\sqrt{s_{NN}} = 200$ GeV are not suppressed but
they scale with $\langle N_{\mathrm{coll}}\rangle$.

\begin{figure}[t]
\begin{center}
\includegraphics[width=0.7\linewidth]{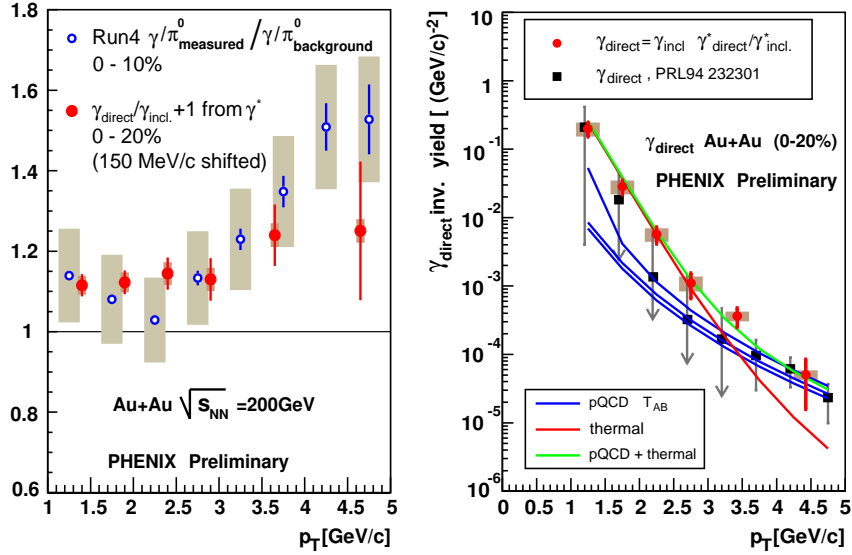}
\end{center}
\vspace*{-10mm}
\caption{\label{fig:photons}
Photon excess (left) and invariant yield of direct photons (right)
as a function of $p_T$ for central Au+Au collisions for two different analysis techniques:
the conventional method (open circles, squares) 
as described in \cite{Adler:2005ig} and a new
method based on virtual photons (closed circles). 
Statistical and systematic errors are indicated separately on each data point by the 
vertical bar and shaded box, respectively. The invariant yield is compared
to various theoretical calculations.}
\end{figure}

At lower $p_T < 4$ GeV the measurement of direct photons is extremely challenging. As the background of
decay photons is less suppressed, the signal to background ratio is small and the systematic uncertainties 
in the extraction of a direct photon signal are large. 
Nevertheless, with the extended Au+Au data set from RHIC Run-4 (2004), it was possible to refine 
the analysis technique - using an improved particle identification 
and strict quality control in run selection - to decrease the 
systematic errors. The result of this analysis is shown in Fig.~\ref{fig:photons} (left) as the
ratio of the measured inclusive photon yield and the expected yield of 
``background'' photons 
from hadronic decays for the $0-10\%$ most central events. 
This $p_T$ region is of particular interest in the search for 
thermal direct photons, which are produced both in the QGP phase and 
in the hadron-gas phase at low $p_T$.
Theoretical predictions~\cite{Turbide:2003si} indicate that at the same temperature the photon 
production rates in a QGP and a hadron gas are very similar but with a small possibility to distinguish
between the two production scenarios in the region $ 1 < p_T < 3$ GeV.

To improve the experimental handle on direct photon production at low $p_T$, 
PHENIX is applying a new analysis technique for heavy ion
experiments that measures real direct photons through the $e^+ e^-$
decay of accompanying very-low-mass virtual direct photons~\cite{QMStefan}.
This was only possible due to the combination of an excellent mass resolution at 
low invariant mass and a low conversion probability in the PHENIX experiment.
The new method allows to improve upon both the signal-to-background 
ratio and the energy resolution compared to the conventional method 
(as described above and in \cite{Adler:2005ig}).

This new method is built on the basic assumption that any source of real photons, 
and of real direct photons in particular, produces virtual photons with very
low invariant mass. The invariant mass distribution of these virtual photons
calculated from the decay $e^+ e^-$ pairs can be described similar to the 
Dalitz decay of $\pi^0$'s. Decay photons can mostly be eliminated by measuring 
the yield of $e^+ e^-$ pairs in an invariant mass region where pairs from the
$\pi^0$ Dalitz decay are largely suppressed due to their limited phase
space. The yield obtained in this invariant mass bin is then normalized to the 
yield obtained in an invariant mass region where the phase space of decay 
photons is unrestricted \cite{QMStefan}. The yield of real direct
photons is then obtained from the measured yield of virtual direct photons
under the assumption that $
{\gamma_{\mathrm direct}^\star}/
{\gamma_{\mathrm incl.}^\star} = 
{\gamma_{\mathrm direct}}/
{\gamma_{\mathrm incl.}}$
Here the yield of real inclusive photons is measured with the EMCal\cite{Adler:2005ig}.

Figure~\ref{fig:photons} (left) shows the photon excess ${\gamma_{\mathrm direct}}/
{\gamma_{\mathrm incl.}} + 1$ compared to the new preliminary result from the 
conventional analysis for the $0-20\%$ most central events.
A significant excess of direct photons at $1 \mbox{GeV}/c < p_T < 5 \mbox{GeV}/c$
beyond systematic errors is observed.
Figure~\ref{fig:photons} (right) shows the deduced invariant yield as a function
of $p_T$ compared to the result from the RHIC Run-2 (2002) \cite{Adler:2005ig}.
Furthermore, the invariant yield is compared to a pQCD calculation~\cite{Jager:2002xm},
a thermal-photon calculation~\cite{d'Enterria:2005cs} and the sum of the two.
The yield lies significantly above the expectation from the NLO pQCD calculation and
is consistent with rates calculated when thermal photon emission is
taken into account. Several similar models can describe the data as well, they all
assume a maximum temperature at the center of 
the fireball $T_{max}$ of 500-600 MeV. Averaging over the entire volume leads to 
smaller values of 300-400 MeV.
This would be the first direct measurement of the initial temperature of the produced
matter if the yield indeed were due to thermal radiation.
However, to draw final conclusions a reference measurement in p+p and d+Au is 
definitely needed. 

\section{J/$\psi$ PRODUCTION}

\begin{figure}[t]
\begin{center}
\includegraphics[width=0.7\linewidth]{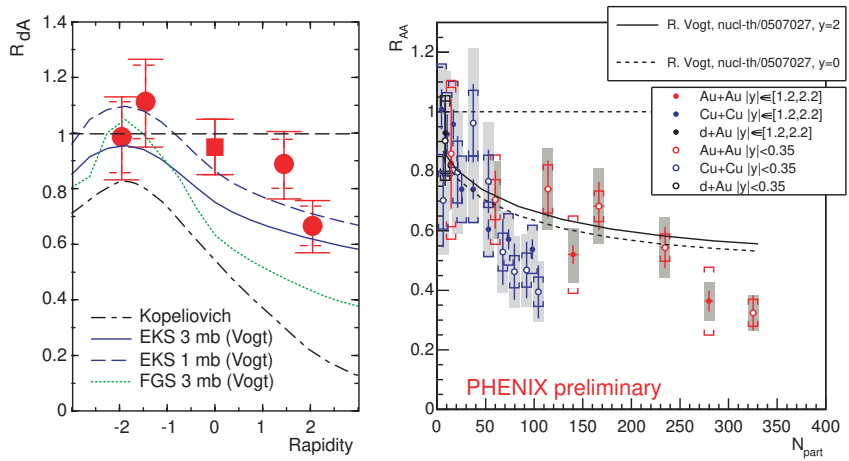}
\end{center}
\vspace*{-10mm}
\caption{\label{fig:jpsi1} J/$\psi$ nuclear modification factor compared to predictions for 
normal nuclear absorption.
Left panel: minimum bias $R_{dA}$ as a function of rapidity. 
Right panel: $R_{AA}$ as a function of the number of participants.
Vertical bars show statistical errors, brackets show point-to-point 
and boxes show global systematic errors.}
\end{figure}

In the quest for understanding heavy quarkonia and especially J/$\psi$ production in heavy ion 
collisions PHENIX now has collected a wealth of new data:
during the RHIC data taking periods Run-3 to Run-5 (2003-2005) 
PHENIX was able to measure J/$\psi$ production
in p+p, d+Au, Cu+Cu and Au+Au collisions at $\sqrt{s_{NN}}=200$~GeV.
This provides access to heavy ion collisions 
and crucial reference systems in the very same experiment. 
Furthermore, PHENIX has the possibility to measure J/$\psi$ production in two decay channels:
J/$\psi\rightarrow e^+e^-$ at mid-rapidity ($|\eta|<0.35$) and 
J/$\psi\rightarrow \mu^+\mu^-$ at forward rapidity ($|\eta|\in[1.2,2.2]$).

To interpret the production mechanisms in heavy ion collisions it is 
indispensable to first understand 
(in p+p and d+Au collisions)
the cold nuclear effects such as gluon
shadowing or saturation, absorption, initial-state energy loss, and the
Cronin effect or $p_T$ broadening~\cite{QMWei}.
Figure~\ref{fig:jpsi1} (left) shows the nuclear modification factor
$R_{dA}$ for J/$\psi$ production in minimum bias d+Au collisions as a function
of rapidity~\cite{ppg038}.
The result is compared to baseline calculations for normal nuclear absorption in combination with a parametrization 
for shadowing effects. The EKS98 calculations~\cite{Vogt2004} assuming an absorption cross section of 
about $\sigma_{abs}$ = 1 mb fit the data best. Models assuming large shadowing effects like FGS~\cite{Vogt2003} 
or the coherence-length model~\cite{KopelivichCoherence} are disfavored. To come to a definite conclusion
an analysis based on higher statistics is desirable.

\begin{figure}[t]
\begin{center}
\begin{tabular}{cc}
\includegraphics[width=75mm]{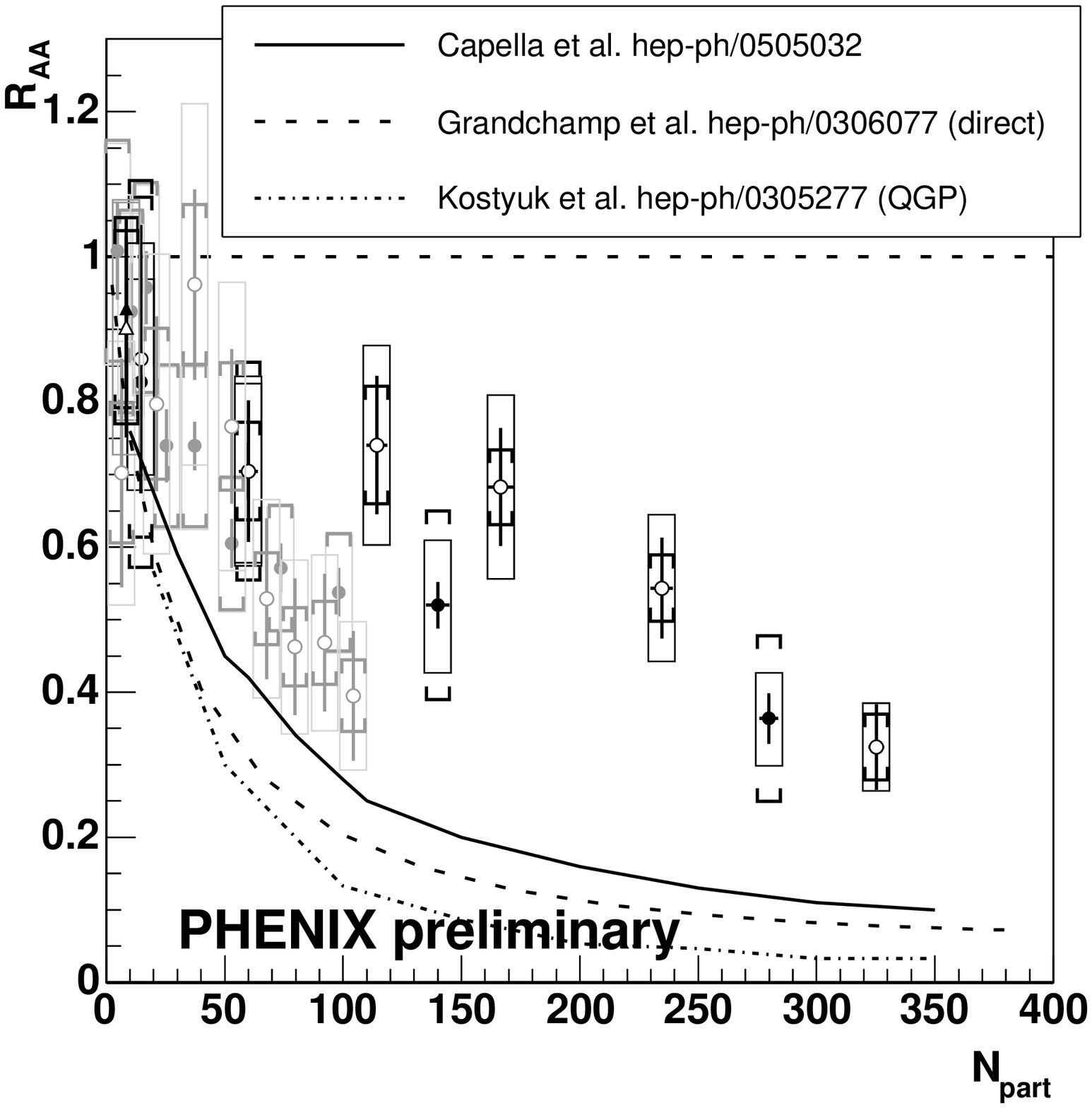}&
\includegraphics[width=75mm]{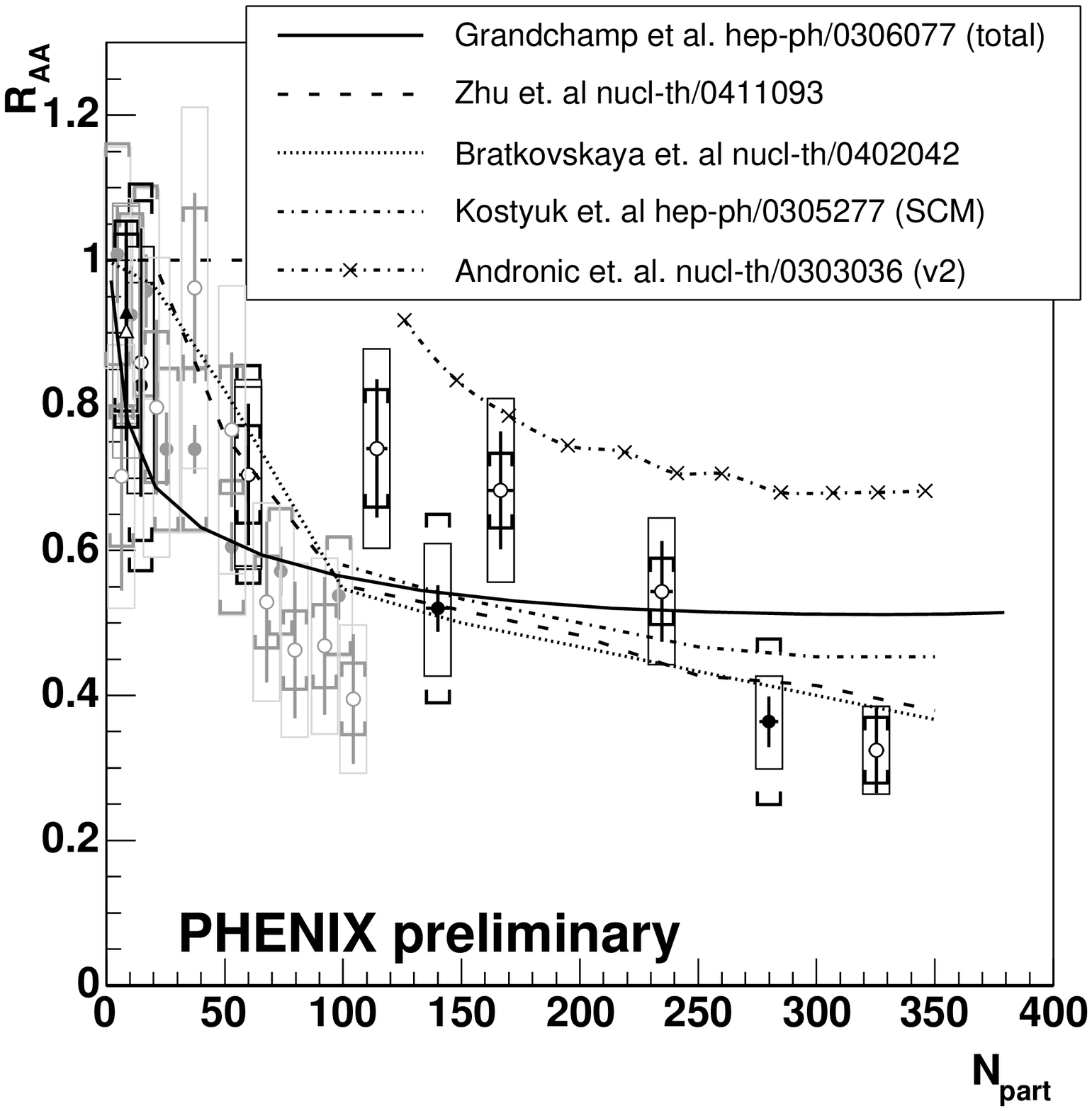}\\
\end{tabular}
\end{center}
\vspace*{-10mm}
\caption{\label{fig:jpsi2}J/$\psi$ nuclear modification factor in d+Au,
 Au+Au and Cu+Cu collisions as a function of the number of participants
 compared to various models of final-state interaction in the medium.
 Data are the same on both panels and on Fig.\ref{fig:jpsi1} (right).
 Vertical bars are statistical errors, brackets are point to point 
 systematics and boxes are global systematics.}
\end{figure}

Figures~\ref{fig:jpsi1} (right) and~\ref{fig:jpsi2} show the nuclear modification factor
$R_{AA}$ for J/$\psi$ production in Au+Au and Cu+Cu collisions as a function of the number of participants
for two different rapidity ranges~\cite{QMHugo}. All data points seem to follow the same general trend.
In the most central collisions a suppression of about a factor 3
relative to binary-scaled p+p collisions is observed.

To interpret the results the data can be compared to various theoretical calculations.
Figure~\ref{fig:jpsi1} (right) shows a comparison with the baseline calculation of the normal-nuclear-absorption model introduced
above~\cite{VOGT} for two different rapidity selections.
The model can describe the data within errors except for the most central data points.
However, in this calculation an absorption cross section of 3~mb in
combination with EKS98 shadowing is used,
which seems to overpredict the suppression seen for the d+Au data in Fig.~\ref{fig:jpsi1} (left). 

Figure~\ref{fig:jpsi2} shows a comparison with two different types of theoretical calculations incorporating additional 
final state interaction in the medium. At SPS energies an {\em anomalous} suppression of J/$\psi$ production relative to binary scaled 
p+p collisions was observed that exceeds the {\em normal} nuclear absorption~\cite{SPS}. The models shown in this comparison
were able to qualitatively reproduce this {\em anomalous} suppression by including color-screening effects in the final state.
A prediction of these models for RHIC energies is shown on the left~\cite{CAPELLA,GRANDCHAMP,KOSTYUK}.
The models shown on the right include either additional quark recombination mechanisms~\cite{GRANDCHAMP,KOSTYUK,ANDRONIC,BRATKOVSKAYA} 
or a detailed J/$\psi$ transport in the medium~\cite{ZHU}. 
One can easily see that the calculations on the left clearly overestimate the J/$\psi$ suppression.
The models on the right however show a much better agreement with the data or they even underestimate the suppression.
Quark recombination models are consistent with the data, although the
recombination component seems to be a little too strong for the most
central collisions.

An important test for quark-recombination mechanisms is the ability to describe the rapidity dependence of J/$\psi$ production
and the evolution of $\langle p^2_T \rangle$ with centrality.
A discussion of new preliminary results from PHENIX on these studies can be found in~\cite{QMHugo}. 
Though some quark recombination models favor a narrow rapidity distribution,
within the currently still large error bars no significant change in the shape of the distribution is observed from 
p+p collisions to the most central Au+Au collisions in these preliminary results.
Given the large errors the interpretation of the J/$\psi$ $\langle p^2_T \rangle$ evolution remains open as well for now. 

\section{JET CORRELATIONS}
\begin{figure}[t]
\begin{center}
\begin{tabular}{cc}
\includegraphics[width=0.48\linewidth]{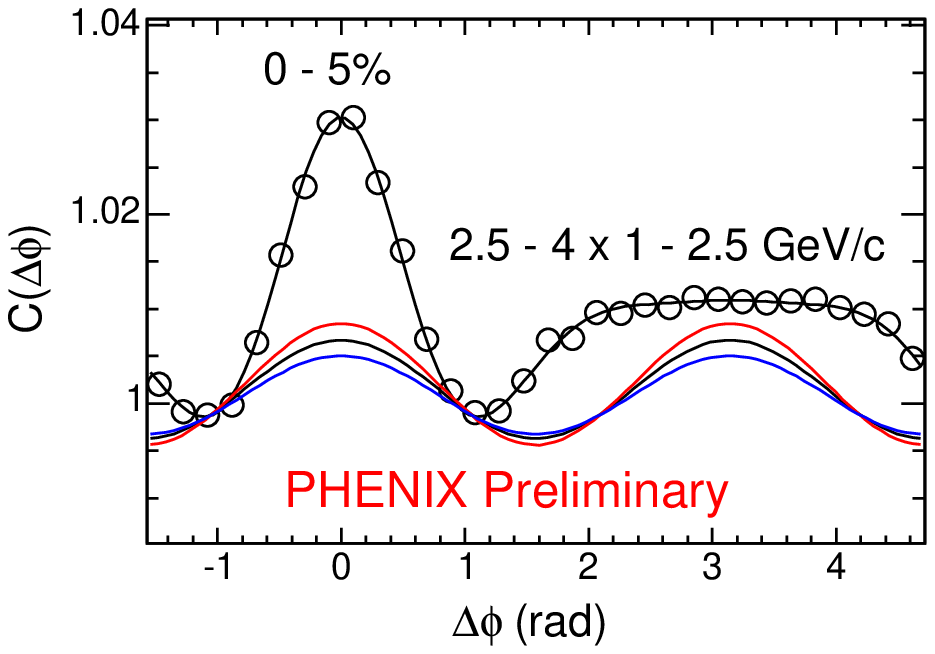}&
\includegraphics[width=0.48\linewidth]{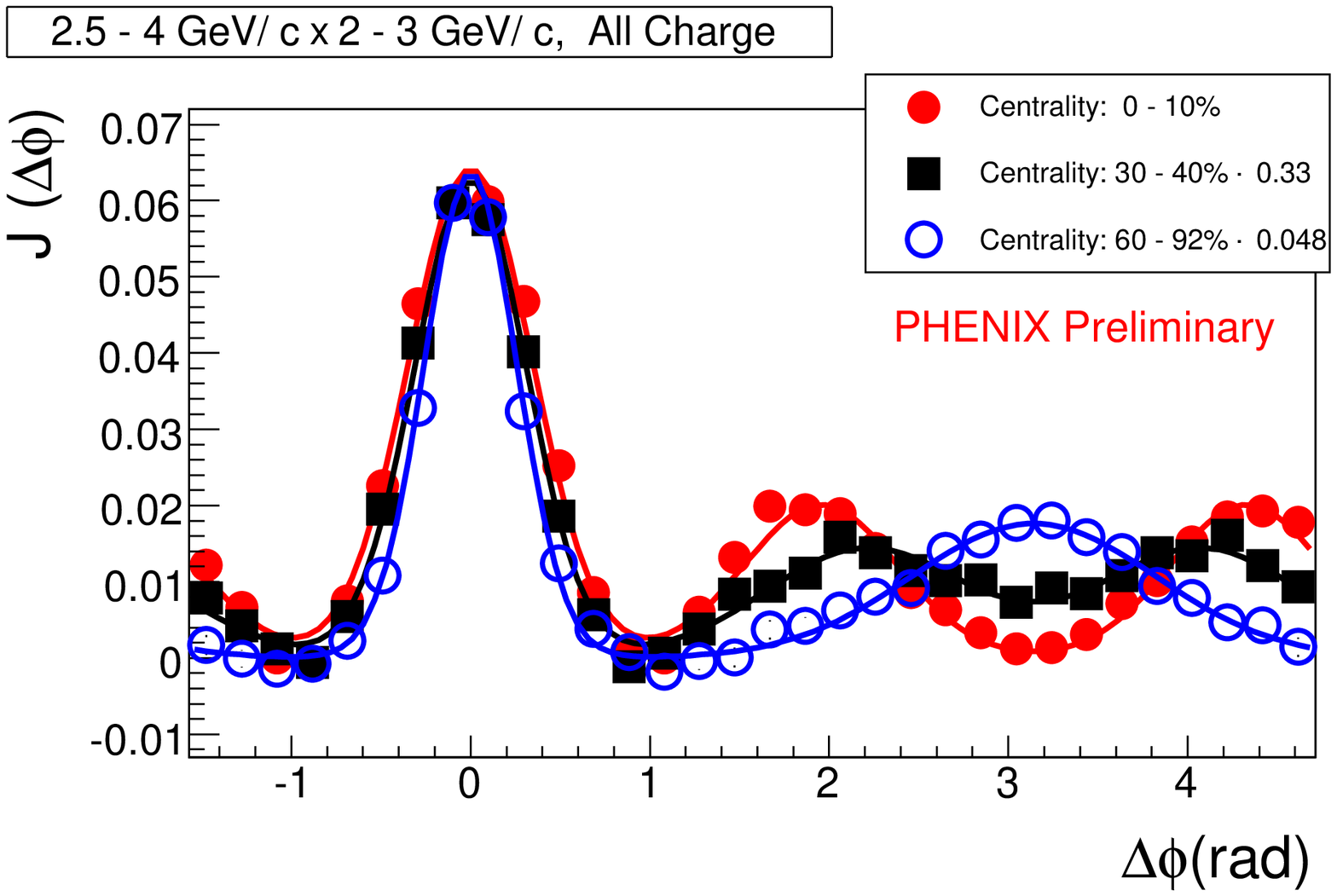}\\
\end{tabular}
\vspace*{-10mm}
\caption{\label{fig:jet1} Left panel: correlation function for the 5\% most central reactions in 
Au+Au collisions at $\sqrt{s_{NN}} = 200$ GeV. The lines indicate the level of flow background 
and its systematic error.
Right panel: correlation function for central, mid-central, and peripheral reactions after subtraction 
of the flow contribution.}
\end{center}
\end{figure}

Two-particle correlations have been proven to be a powerful tool in the understanding of already the
first Au+Au results at $\sqrt{s_{NN}} = 200$ GeV from Run-2 at RHIC (2002). It was 
shown that the away-side jet in central Au+Au collisions at moderately high $p_T$ for both the trigger 
and the associated hadrons is largely
extinguished~\cite{StarAwayJet}. Furthermore, it was found recently that at intermediate
trigger $p_T$ the shape of the away-side jet is modified in a complex way~\cite{PhenixAwayJet}. 
With the extended Run-4 Au+Au data set (2004) it is now possible to study particle 
correlations in much more detail to distinguish different contributions to the 
correlation function.

To describe the two-particle correlations a correlation function is constructed as 
\begin{equation}\label{eqn:corr}
C\left(\Delta\phi\right) \propto
\frac{dN_{real}/d\Delta\phi}{dN_{mix}/d\Delta\phi}
\end{equation}
where $dN_{real}/d\Delta\phi$ describes the pair distribution in a real event and
$dN_{mix}/d\Delta\phi$ describes the mixed-event distributions where
each particle of the pair is chosen from a random event.  
$dN_{mix}/d\Delta\phi$ corrects for the combinatorial background and the 
geometrical pair acceptance of the detector.

In heavy ion collisions the main contributors to the correlation function are
jets and elliptic-flow. A typical correlation function for 
central Au+Au collisions at $\sqrt{s_{NN}} = 200$ GeV is shown in 
Fig.~\ref{fig:jet1} (left). The correlation function
exhibits a peak at $\Delta\phi$=0 (near side) and a broad peak at
$\Delta\phi$=$\pi$ (away-side) originating from jets.
They represent correlations within the same jet and between back-to-back jets, respectively. 
Elliptic flow causes a harmonic modulation of the background.

PHENIX has developed various methods to determine the contribution of
flow effects to the correlation function and to subtract it~\cite{QMNathan,Jia:2005ab,QMAjit}. 
All correlation functions that are discussed in the following show the correlation {\it after}
subtraction of the flow component.

Fig.~\ref{fig:jet1} (right) shows an example of such a correlation function, $J(\Delta\phi$), for hadrons 
with a trigger particle in the intermediate $p_T$ range ($2.5 < p_T < 4.0$ GeV/c) and the associated 
particle with $2.0 < p_T < 3.0$ GeV/c for different centrality selections.
In peripheral collisions the clear structure of near- and away-side jet peaks can be seen.
Moving to more central events, the away-side peak gets broader and develops a ``dip'' 
at $\Delta\phi$=$\pi$ - a feature that can already be seen in Fig.~\ref{fig:jet1} (left)
before flow subtraction.

\begin{figure}[t]
\begin{center}
\begin{tabular}{cc}
\includegraphics[width=0.48\linewidth]{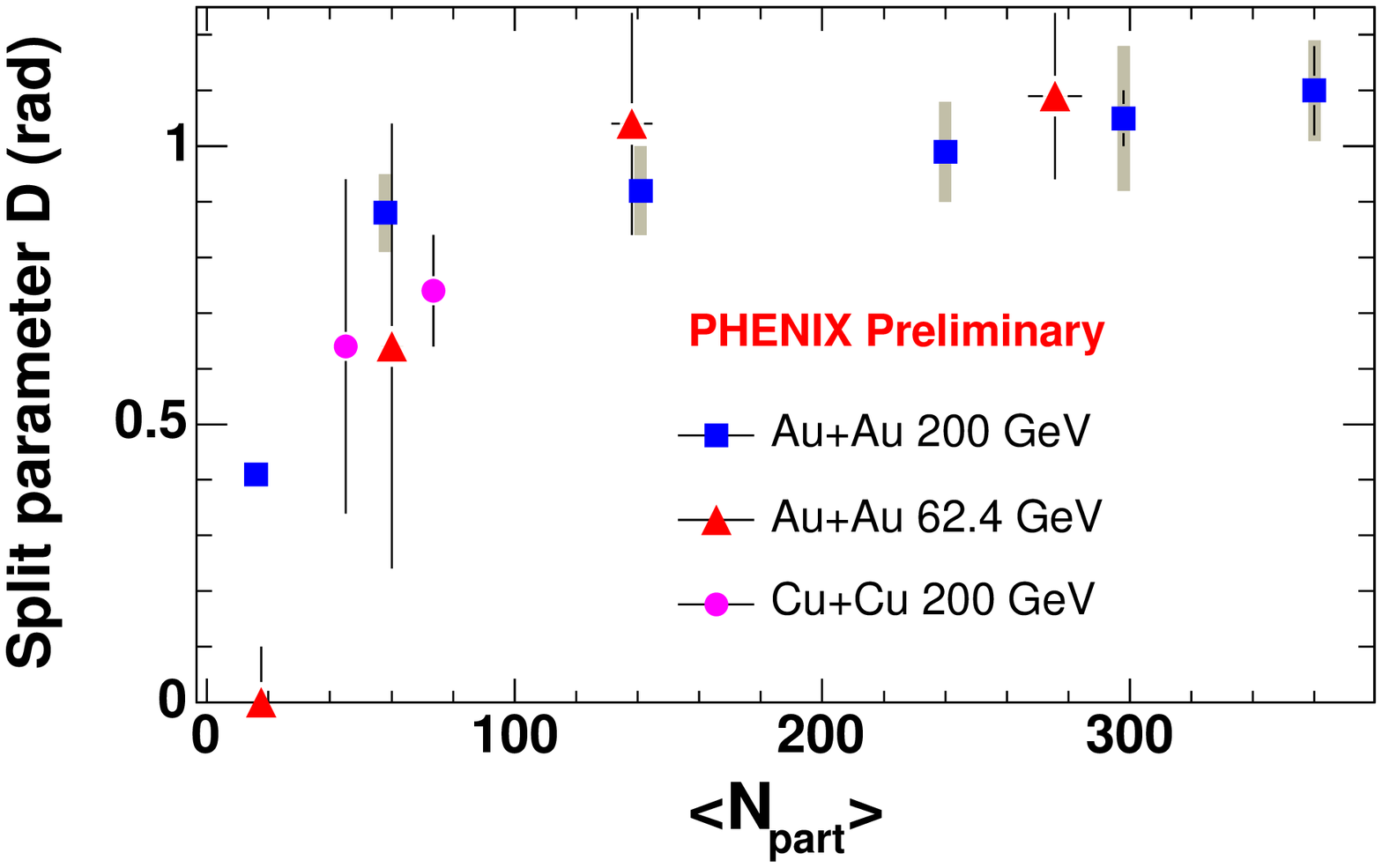}&
\includegraphics[width=0.48\linewidth]{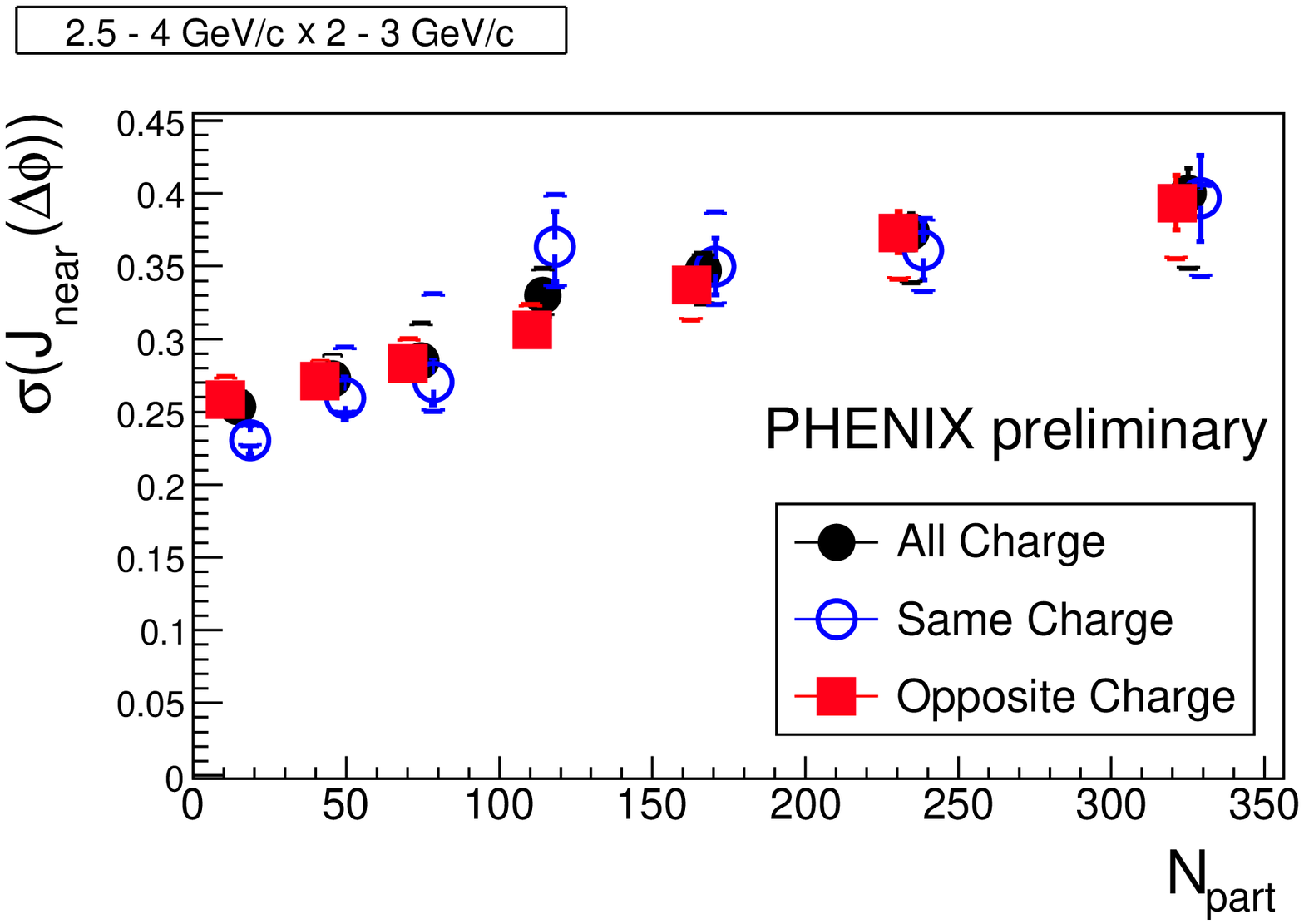}\\
\end{tabular}
\vspace*{-10mm}
\caption{\label{fig:jet2} 
Left panel: the away-side splitting parameter ``D'' as a function of $N_{part}$.
(right): near side width as a function of centrality for same charged pairs, opposite charged pairs and all pairs.} 
\end{center}
\end{figure}

PHENIX has performed a systematic study of the jet shape at intermediate $p_T$. The result is
shown in Fig.~\ref{fig:jet2}. 
To characterize the splitting of the away-side peak a splitting parameter D is introduced~\cite{QMNathan}.
It is obtained by a double Gauss fit on the away side. D describes the distance between the two
mean values of the Gauss fits. The dependence of the splitting parameter D on the centrality of the collision 
is shown in Fig.~\ref{fig:jet2} (left). It can be seen that D evolves
smoothly with increasing centrality
and seems to follow the same trend for all reaction systems.
The splitting of the away side happens already in semi-peripheral events; for more central events D seems to be 
almost constant.

The observation of a splitting pattern in the away-side jet has already met a lot of interest in the 
theory community~\cite{Koch:2005sx,Armesto:2004vz,Casalderrey-Solana:2004qm}. As the standard energy 
loss will broaden the jet but is unable to produce a dip or a flat jet structure in the away side,
alternative models have been developed including Cherenkov-like effects or a Mach cone/shock wave mechanism.

In addition to the modification of the away-side jet a moderate broadening of the near side jet width with increasing
centrality at intermediate trigger $p_T$ is shown in Fig.~\ref{fig:jet2} (right). This could be consequence
of the strong interaction of jets with the medium.

\begin{figure}[t]
\begin{center}
\includegraphics[width=\linewidth]{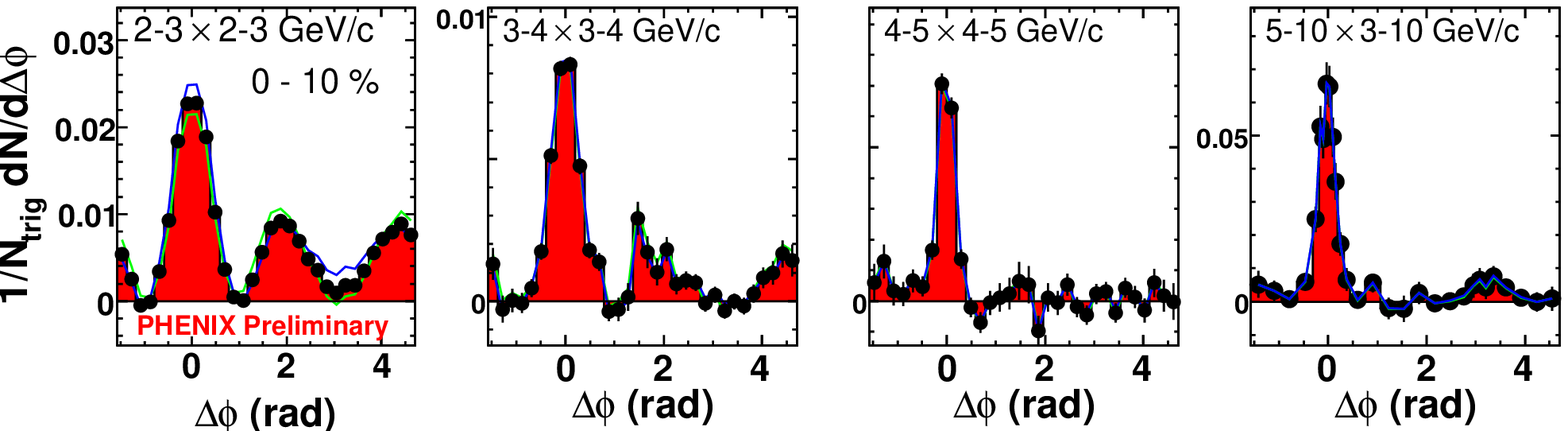}
\includegraphics[width=\linewidth]{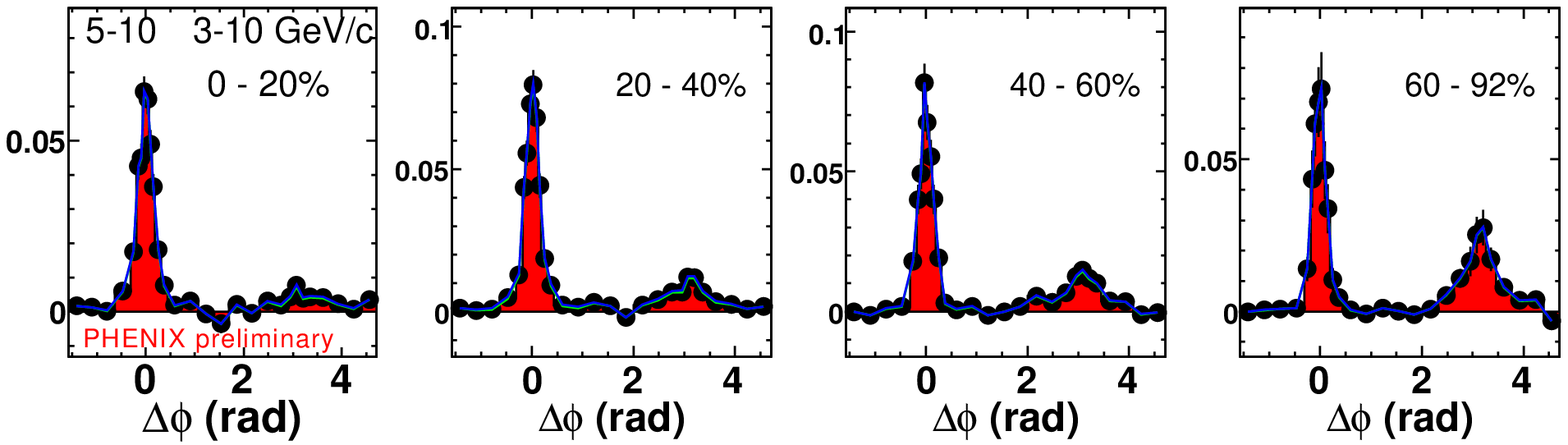}
\end{center}
\vspace*{-10mm}
\caption{\label{fig:jet3} Top: per-trigger yield of hadron pairs 
for different $p_T$ selections in the 10\% most central 
Au+Au collisions at $\sqrt{s_{NN}} = 200$ GeV. The $p_T$ range is given in the plot
(trigger $p_T \times$ associated $p_T$).
Bottom: centrality dependence of the per-trigger yield at high trigger
$p_T$.  The trigger and associated $p_T$ ranges are the same on all
four panels.}
\end{figure}

With the extended statistics of the Run-4 Au+Au data (2004) it now becomes possible to study jet
correlations at high $p_T$ as well. Particle production at high $p_T$ is dominated by 
parton-parton interactions with high momentum transfer. 
To get a complete picture of the interaction of the jet with the medium, it is important
to understand the correlations at high $p_T$ as well.
Furthermore, the evolution of jet correlation patterns from intermediate to high $p_T$
can help to understand the complicated medium effects just described for intermediate $p_T$.

Figure~\ref{fig:jet3} (top) shows the evolution of the per trigger yield for hadron pairs for the
10\% most central Au+Au collisions at $\sqrt{s_{NN}} = 200$ GeV with increasing $p_T$.
Here the $p_T$ of the trigger particle is chosen similar to the $p_T$ of the associated particle.
It can be seen that the ``dip'' structure in the away-side peak at intermediate $p_T$ becomes weaker
as the $p_T$ of the trigger particle is increased. At ($4-5\times 3-5$) GeV/c the away-side jet 
appears to be flat within errors. In the highest $p_T$ selection in Fig.~\ref{fig:jet3} (top)
an away-side peak seems to re-emerge at $\Delta\phi$=$\pi$.

For this $p_T$ selection the evolution with the centrality of the collision is shown in 
Fig.~\ref{fig:jet3} (bottom). The away-side peak is visible in all centrality selections
but the suppression is larger for more central collisions. 
Given the current statistics it is still difficult to see whether the away-side
peak is broadened in central collisions.

\begin{figure}[t]
\begin{center}
\begin{tabular}{ccc}
\includegraphics[width=0.3\linewidth]{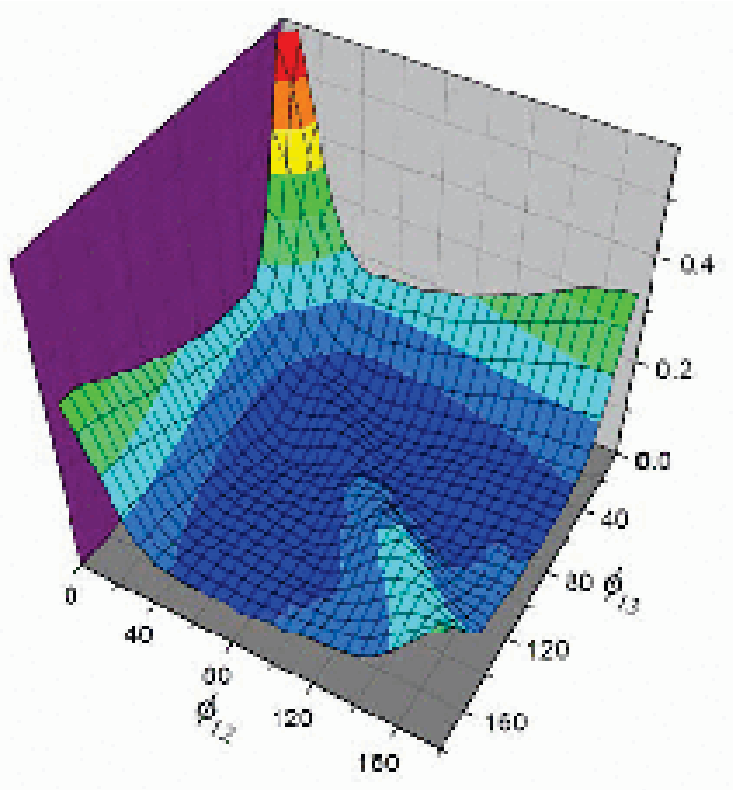}&
\includegraphics[width=0.3\linewidth]{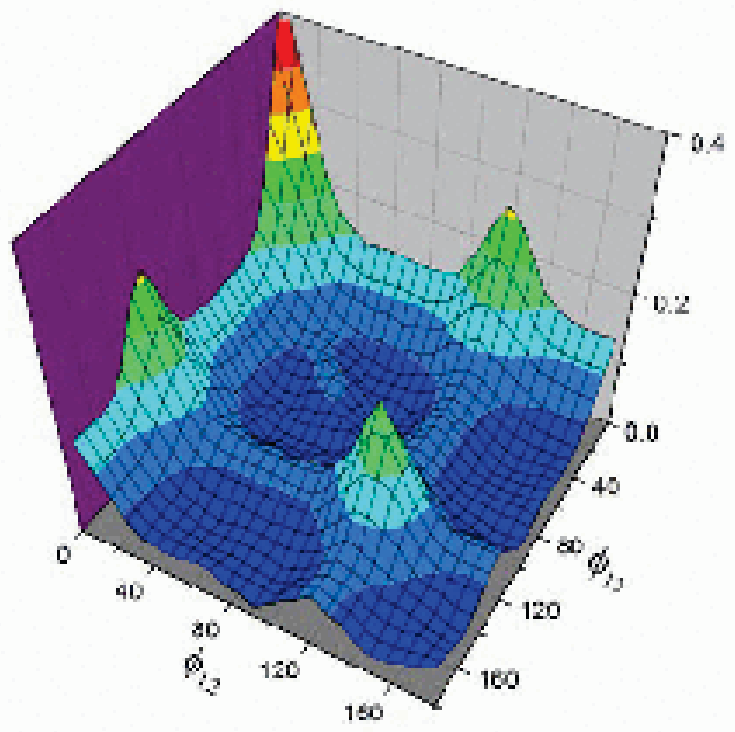}&
\includegraphics[width=0.3\linewidth]{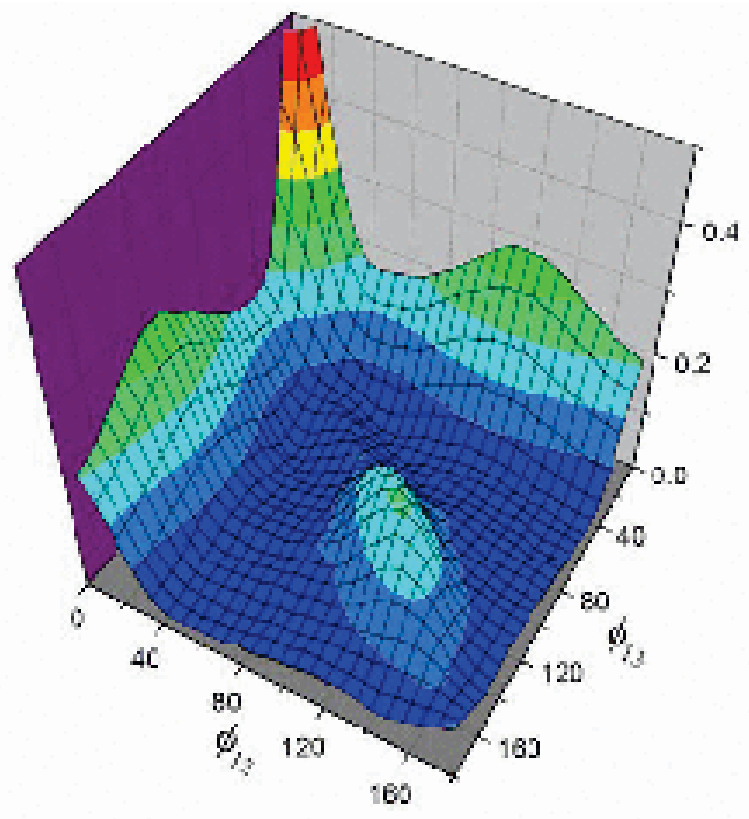}\\
\includegraphics[width=0.3\linewidth]{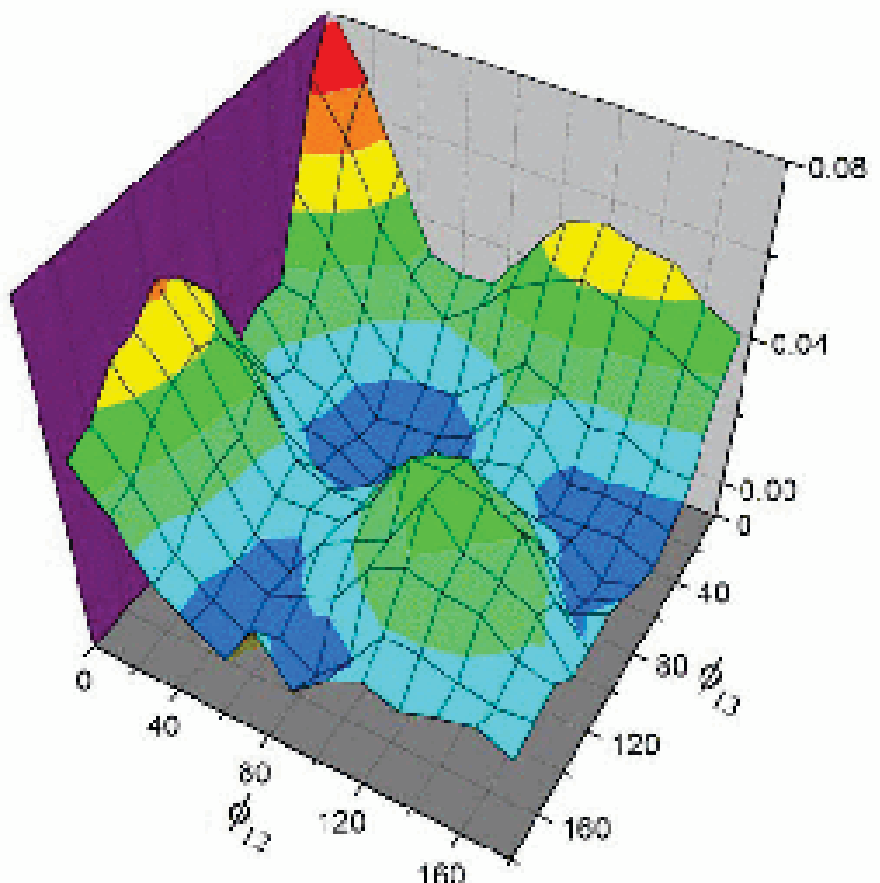}&
\includegraphics[width=0.3\linewidth]{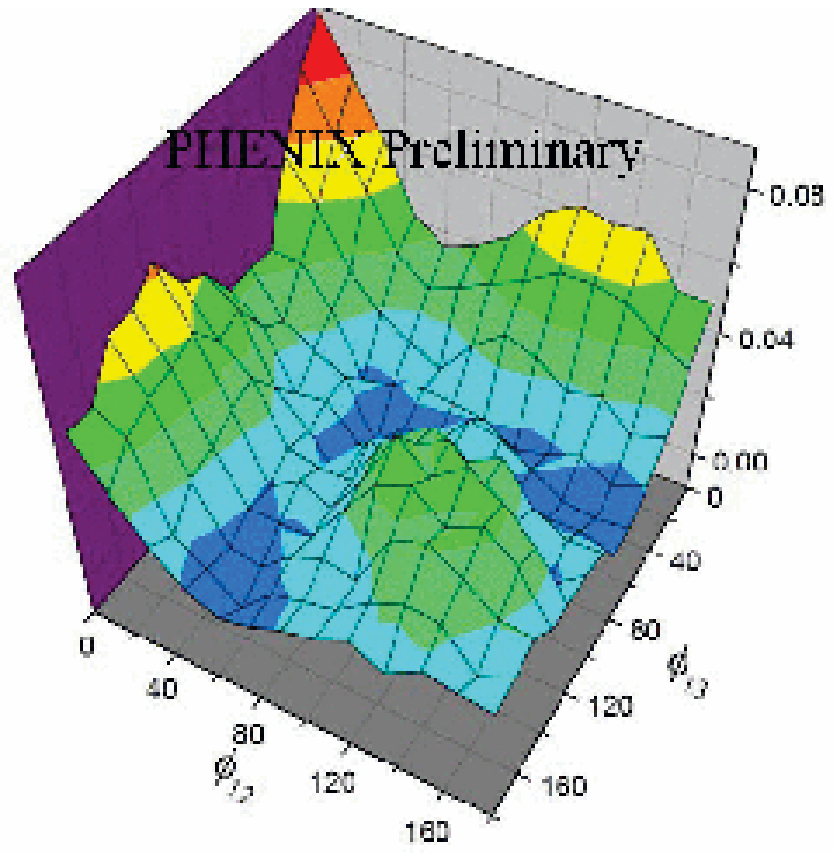}&
\includegraphics[width=0.3\linewidth]{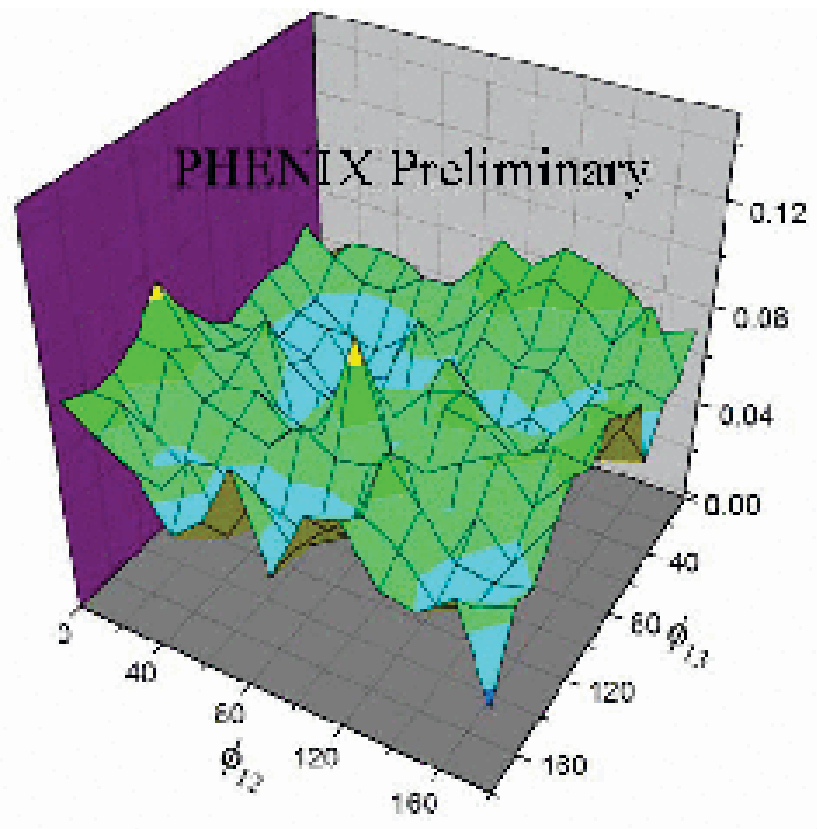}\\
\end{tabular}
\end{center}
\caption{\label{jet3d}Three-particle correlations. Top: simulations for three different scenarios. 
From left to right: normal jets, Cherenkov or conical jets, and ``bent'' jets.
Bottom: data for three different particle selections. From left to right:
hadron-hadron-hadron, hadron-meson-meson and hadron-baryon-baryon correlations.}
\end{figure}

Finally first results on three-particle correlations are presented.
To construct the correlation function in this case the relative angles $\Delta\phi_{1,2}$ and $\Delta\phi_{1,3}$ between a
trigger particle with $2.5 < p_T < 4.0$ GeV/c and two associated particles with $1.0 < p_T < 2.5$ GeV/c are calculated.
The distribution of $\Delta\phi_{1,2}$ vs. $\Delta\phi_{1,3}$ constitutes the correlation surface.
Mixed events were created by combining the high-$p_T$ trigger particle of one event with associated particles from 
two other events. Therefore the correlation functions contain by construction contributions from both two- and three-particle
correlations.

To illustrate the motivation a selection of simulated correlation functions is shown in Fig.~\ref{jet3d} (top).  
The figure shows simulations for three different scenarios~\cite{QMAjit}:
(i) {\it normal jets}, where the away-side jet axis is aligned with the leading jet axis with a certain smearing;
(ii) {\it Cherenkov or conical jets}, where the leading and away-side jet axes are aligned
 but the fragmentation is confined to a hollow cone with an opening angle of $60^o$ but with no smearing on the jet itself.
(iii) {\it ``bent'' jets}, where the away-side jet axis is misaligned by $60^o$, together with the same smearing as for normal jets;
In this simulation the three different scenarios show clear distinguishing features.

The actual measurements of $C(\Delta\phi_{1,2},\Delta\phi_{1,3}) = N_{real}/N_{mix}$ 
are shown in Fig.~\ref{jet3d} (bottom) for the 10-20{\%} most central events.
Here the flow component is subtracted by choosing the orientation of the trigger
particle relative to the reaction plane of the event such that the flow component of the 
correlation vanishes.
In the figure the correlation function for three different particle selections is shown:
a high-$p_T$ trigger hadron is combined with two other hadrons, two mesons, or two baryons. 
The data show a strong dependence on the flavor of the associated particle:
baryons and mesons as associated particles show a very different correlation pattern.
Furthermore the correlation pattern clearly does not follow the expected pattern for a normal jet
from the simulation. This is additional evidence for the strong modification of the away-side jet.

To come to a final conclusion on the nature of the modification based on three-particle 
correlations additional detailed studies are necessary.

\section{CONCLUSIONS}
Only a small fraction of the vast amount of new results from the PHENIX collaboration presented at QM05
could be selected for the overview talks. PHENIX presented 18 parallel presentations and 52 posters at
the conference. Since the recent review in the PHENIX {\it White Paper}~\cite{PHENIX_WP} exciting new results
on direct photon production at low $p_T$, J/$\psi$ production, and the modification of jets have been 
added.  We presented first results on a new technique in heavy ion physics to extract direct photons 
at low $p_T$ based on a $e^+ e^-$ pair analysis. With this new method a significant direct photon signal 
can be measured at $1 \mbox{GeV}/c < p_T < 5 \mbox{GeV}/c$. The yield lies above NLO pQCD
expectations and is consistent with rates calculated when thermal photon emission is taken into account.

PHENIX collected a wealth of new data on J/$\psi$ production in Au+Au and Cu+Cu collisions at various
energies and rapidities. In the most central collisions a suppression by a factor of 3 is observed in comparison
to binary scaled p+p collisions. 
Models implementing color screening effects in addition to quark recombination
are consistent with the centrality dependence of the data, although the
recombination component seems to be a little too strong for the most central collisions.

With the extended statistics of the RHIC 2004 Au+Au run it was possible to improve the understanding of
systematics in the correlations of particle jets: at intermediate trigger $p_T$ a clear signal of a
broadening of the away-side jet with increasing centrality is observed. Simultaneously a ``dip'' structure
is developing in the away-side jet with increasing centrality.
At high trigger $p_T$ the away-side peak is more and more suppressed as the collisions become more central.
However, even at the highest trigger $p_T$ available the
characteristic jet shape is still visible at the away side, even in the
most central collisions.

\end{document}